\documentclass[%
 reprint,
superscriptaddress,
 amsmath,amssymb,
 aps,
]{revtex4-1}

\usepackage{graphicx}
\usepackage{dcolumn}
\usepackage{bm}
\usepackage{color}
\usepackage{float}

\begin{document}

\preprint{APS/123-QED}

\title{Self-assembly of binary solutions to complex structures}

\author{Alberto Scacchi}\email{alberto.scacchi@aalto.fi}
\affiliation{Department of Chemistry and Materials Science, Aalto University, P.O. Box 16100, FI-00076 Aalto, Finland}
\affiliation{Department of Applied Physics, Aalto University, P.O. Box 11000, FI-00076 Aalto, Finland}
\author{Maria Sammalkorpi}
\affiliation{Department of Chemistry and Materials Science, Aalto University, P.O. Box 16100, FI-00076 Aalto, Finland}
\affiliation{Department of Bioproducts and Biosystems, Aalto University, P.O. Box 16100, FI-00076 Aalto, Finland}
\author{Tapio Ala-Nissila} 
 \affiliation{Quantum Technology Finland Center of Excellence and Department of Applied Physics, Aalto University, P.O. Box 11000, FI-00076 Aalto, Finland}
 \affiliation{Interdisciplinary Centre for Mathematical Modelling and Department of Mathematical Sciences, Loughborough University, Loughborough, Leicestershire LE11 3TU, United Kingdom}

\date{\today}

\begin{abstract}
Self-assembly in natural and synthetic molecular systems can
create complex aggregates or materials whose properties and functionalities rise from their internal structure and molecular arrangement. The key microscopic features that control such assemblies remain poorly understood, nevertheless.
Using classical density functional theory we demonstrate how the intrinsic length scales and their interplay in terms of interspecies molecular interactions can be used to tune soft matter self-assembly. 
We apply our strategy to two different soft binary mixtures to create guidelines for tuning intermolecular interactions that lead to transitions from fully miscible, liquid-like uniform state to formation of simple and core-shell aggregates, and mixed aggregate structures. Furthermore, we demonstrate how the interspecies interactions and system composition can be used to control concentration gradients of component species within these assemblies. The insight generated by this work contributes towards understanding and controlling soft multi-component self-assembly systems. Additionally, our results aid in understanding complex biological assemblies and their function and provide tools to engineer molecular interactions in order to control polymeric and protein-based materials, pharmaceutical formulations, and nanoparticle assemblies. 
\end{abstract}

\maketitle


\section{Introduction}

In self-assembly of molecular systems, a disordered solution spontaneously organizes to form ordered structures or assembly patterns from its components.  Examples of natural and synthetic materials where self-assembly, the resulting complex structures, and their advanced functional properties are central~\cite{whitesides2002beyond} include molecular crystals \cite{wang2020,desiraju1989crystal}, colloids \cite{evans1999thecolloidal}, lipid membranes \cite{jones1995micelles}, polymer assemblies \cite{lu2020,hu2014,stuart2010emerging}, self-assembled monolayers \cite{kumar1995patterned,smith2004} and quasicrystalline structures \cite{lifshitz2007soft}. Basic biological cell function relies on self-assembly phenomena such as spontaneous condensation \cite{banani2017biomolecular} or lipid raft formation \cite{Sezgin2017}. Also technologically, self-assembling soft materials are receiving a great deal of attention in a variety of applications including drug delivery \cite{yadav2020,branco2009self,kataoka2012block} and anticancer therapy \cite{huang2020,oerlemans2010polymeric}, nanotechnology photonics \cite{ma2019,wang2020,yang2018,shaikh2021},  polymer \cite{shaikh2021,yang2018} and other electronics \cite{yuan2009facile}, as well as biointerfaces and bioengineering \cite{lyons2020,zhang2002design}.

At the level of thermodynamics the properties of molecular self-assemblies in surfactant, biomolecular, and polymer solutions, such as micellization, emulsion formation, or self-assembling monolayers and bilayers can be explained at the level of interplay between water entropy, the relevant surface tensions and corresponding free energies \cite{nagarajan1991theory,birdi2015handbook,tadros2016emulsions}. However, understanding self-assembly systems involving multiple components, especially the mechanisms giving rise to the complex assembly characteristics in them, requires more advanced theoretical efforts.
To this end, in this manuscript, we present a general theoretical framework based on classical density functional theory (DFT) \cite{hansen_mcdonald, evans_79, lutsko2010recent} with competing molecular interactions. We demonstrate how such cases can lead to molecular solutions transitioning from uniform mixtures or simple particle formation to spontaneous ordered, structurally more complex assemblies. Qualitatively similar molecular-level aggregates range from core-shell molecular assemblies (micelles or polymer particles) to assemblies with internal composition gradients or layered structures (onion-like). Such assemblies are widely present in surfactant and biosystems (micelles, vesicles, emulsion type formulations, biomolecular droplets)~\cite{sorrenti2013,banani2017biomolecular,sheth2020multiple} but also readily formed by polymer assemblies \cite{lu2020, hu2014, stuart2010emerging,shaikh2021}. The DFT work here is based on effective potentials typically used to describe soft, intermixing species. Understanding the spontaneous assembly transitions and controlling the composition gradients, for example, hydrophobic-hydrophilic gradients, is particularly important for designing solubilization and partitioning systems e.g. for separation and extraction \cite{mortada2021} or drug delivery systems~\cite{de2012polymeric, ghosh2012core, haag2004supramolecular, kakizawa2002block}. The self-assembling micro-environments also form synthetic biology artificial cells \cite{trantidou2017,che2016} and confined catalysis systems \cite{York2017,Gaitzsch2016}.  
%

In DFT, approximate free energy functionals are devised for particular model Hamiltonian. This enables extracting an inhomogeneous average density of the different components in the thermodynamic equilibrium based on energy minimization principles
\cite{hansen_mcdonald, evans_79, lutsko2010recent}. 
In the last few years, DFT has been successfully applied to multi-component systems to describe various other forms of self-assembly, such as superlattice structures \cite{somerville2018density}, quasicrystals~\cite{ratliff2019wave, scacchi2020quasicrystal, subramanian2020density} and general multiple length scale self-assembly~\cite{scacchi2021self}. In this approach, the linear dispersion relation~\cite{archer2012solidification, archer2014solidification, archer2016generation, dispersion_relation, walters2018structural, ratliff2019wave}, characterizing the growth or decay of density modulations in the liquid state, provides the key quantity used to establish guidelines to generate appropriate interparticle potentials. The importance of the linear dispersion relation in constructing the effective interactions is also emphasized by its recent application to accurately capture non-equilibrium phenomena such as laning instability~\cite{scacchi2017dynamical} and to control the macroscopic structures of systems under external shear~\cite{scacchi2020sensitive}.
\section{Theory}
\label{sec:DFT}
Using DFT, we can obtain the equilibrium average density distribution of a binary mixture suspended in a (implicit) background fluid. In other words, our model captures the interactions of regions of two different chemical characteristics in a solution to predict how molecular species, corresponding to those chemical characteristics, would assemble and organize. Although these interaction characteristics emerge from the molecular level, we emphasize that the ``particles" in our model implicitly include interactions rising due to their solution environment. These include e.g. contributions arising from solvent quality for each of the components.

The total grand potential of a system composed by two distinct types of particles is described by
\begin{equation}
\Omega[\rho_1,\rho_2]=\mathcal{F}[\rho_1,\rho_2]+\sum_{i=1,2}\int d\textbf{r} \left[V_i^{\rm ext}(\textbf{r})-\mu_i\right]\rho_i(\textbf{r}),\label{grand_canonical}
\end{equation}
where $\mathcal{F}$ is the intrinsic Helmholtz free energy functional, $V_i^{\rm ext}(\textbf{r})$ is the one-body external potential acting on species $i$, $\mu_i$ the corresponding chemical potential and $\rho_i(\textbf{r})$ the average (inhomogeneous) one body number density. The intrinsic Helmholtz free energy can be written as
\begin{equation}
\mathcal{F}[\rho_1,\rho_2]=\mathcal{F}^{\rm id}[\rho_1,\rho_2]+\mathcal{F}^{\rm exc}[\rho_1,\rho_2],\label{free_energy}
\end{equation}
where the first term is the ideal gas contribution. The latter functional has the known form
\begin{equation}
\mathcal{F}^{id}[\rho_1,\rho_2]=k_{\rm B} T \sum_{i=1,2}\int d\textbf{r} \rho_i(\textbf{r})\left[\ln\left(\Lambda_i^d  \rho_i(\textbf{r})\right)-1\right],\label{ideal}
\end{equation}
where $\Lambda_i$ is the thermal de Broglie wavelength and $d$ is the dimensionality of the system (here we focus on $d=2$ for simplicity, but our findings are expected to be valid in $d=3$ as well, where additional configurations are to be expected). The second term in Eq. (\ref{free_energy}) is the excess Helmholtz free energy functional, which describes the interactions between the particles. Following Ramakrishnan and Yussouff \cite{ramakrishnan1979first}, we opt for an expansion of the latter functional around the homogeneous fluid states $\rho_{i}^0$, $i=1,2$, in a functional Taylor expansion and truncate at second order, leaving us with
\begin{equation}
\begin{split}
\mathcal{F}^{\rm exc}&[\rho_1,\rho_2]=\mathcal{F}^{\rm exc}[\rho_{1}^0,\rho_{2}^0]+\sum_{i=1,2}\int d\textbf{r}\mu_i^{ex}\delta\rho_i(\textbf{r})\\&-\frac{1}{2\beta}\sum_{\substack{i=1,2 \\ j=1,2}}\int d\textbf{r}\int d\textbf{r}'\delta\rho_i(\textbf{r})c_{ij}(\mid \textbf{r} -\textbf{r}'\mid)\delta\rho_j(\textbf{r}'),\label{excess}
\end{split}
\end{equation}
where $\delta\rho_i(\textbf{r}) = \rho_i(\textbf{r}) - \rho_{i}^0$ and $\mu_i^{\rm ex} = \mu_i - k_{\rm B} T \ln(\rho_{i}^0\Lambda_i)$ are the excess chemical potential, $c_{ij}(\textbf{r})$ are the direct correlation functions, and $\beta=1/k_BT$. There are different ways to approximate the latter functions. The choice of the approximation depends on the type of interactions that one considers for the model. For this study, we choose the random phase approximation (RPA), which corresponds to $c_{ij}(\textbf{r})=-\beta \phi_{ij}(\textbf{r})$, where $\phi_{ij}(\textbf{r})$ are effective pair potentials~\cite{likos2001effective}. This approximation has proven to be very efficient when dealing with soft interactions in polymer, surfactant, and other macromolecular systems including dendrimers or star polymers \cite{louis2000mean, likos2001criterion, archer2002binary, archer2015soft,scacchi2021self}, and even for describing structure and dynamics of cells in living tissues and tumors~\cite{al2018dynamical}. The accuracy and the limitations of this approximation for systems relevant to this work are discussed in detail in e.g., Refs.~\cite{archer2014solidification, archer2015soft, archer2002binary}. However, if more accurate approximations were needed, one could for example use the so-called hypernetted chain (HNC) Ornstein-Zernike integral equation theory \cite{hansen_mcdonald} to obtain the direct correlation functions $c_{ij}(\textbf{r})$ in Eq.~(\ref{excess}). Such approach has recently been applied to e.g. describe nanoparticles forming quasicrystals \cite{scacchi2020quasicrystal}. 

The equilibrium average density profiles $\rho_i(\textbf{r})$ are those which minimise the functional of the grand potential $\Omega[\rho_1,\rho_2]$. Therefore, they also satisfy the coupled Euler-Lagrange equations
\begin{equation}
\frac{\delta \Omega[\rho_1,\rho_2]}{\delta\rho_i}=0,
\end{equation}
for $i=1,2$. For bulk systems, i.e. for $V_i^{\rm ext}(\textbf{r}) \equiv 0$ for $i = 1, 2$ in Eq. (\ref{grand_canonical}), this is equivalent to solving the following coupled equations for $\rho_1(\textbf{r})$ and $\rho_2(\textbf{r})$
\begin{equation}\label{solver}
\ln\left(\frac{\rho_i(\textbf{r})}{\rho_i^0}\right)-\beta\sum_{j=1,2}\int d\textbf{r}'c_{ij}(\mid \textbf{r} -\textbf{r}'\mid)\delta\rho_j(\textbf{r}')=0.
\end{equation}
The results presented hereafter are obtained solving the latter equation under periodic boundary conditions. Next, we discuss the linear dispersion relation and the partial liquid structure factors. These quantities provide important guidelines with respect to the length scales and structures that are expected to be found in the inhomogeneous state~\cite{scacchi2020quasicrystal}. They can also be used to find appropriate model parameters. The Ornstein-Zernike equation for binary mixtures and different closures are also briefly presented.
\subsection{The dispersion relation}
The two branches of the dispersion relation for a binary (isotropic) mixture are \cite{dispersion_relation}
\begin{equation}\label{dispersion_relation_eq}
\omega_{\pm}(k)=\frac{1}{2}\textrm{Tr}(\textbf{M}\textbf{E})\pm \sqrt{\frac{1}{4}\textrm{Tr}(\textbf{M}\textbf{E})^2-\textrm{det}(\textbf{M}\textbf{E})},
\end{equation}
where the mobility matrix $\textbf{M}$ and the energy matrix $\textbf{E}$ are defined as
\begin{equation}
\textbf{M}=-k^2\begin{pmatrix}
D_1\rho_{1}^0 & 0\\
0 & D_2\rho_{2}^0
\end{pmatrix},
\end{equation}
and
\begin{equation}
\textbf{E}=\begin{pmatrix}
\left[\frac{1}{\rho_{1}^0}-\hat{c}_{11}(k)\right] & -\hat{c}_{12}(k)\\
-\hat{c}_{21}(k) & \left[\frac{1}{\rho_{2}^0}-\hat{c}_{22}(k)\right]
\end{pmatrix},
\end{equation}
respectively. In these, $\hat{c}_{ij}(k)$ are the Fourier transform of the direct correlation functions $c_{ij}(r)$ and $D_{i}$ is the diffusion coefficient of species $i$.  If $\omega_{\pm} < 0$, for all wave numbers $k=|\textbf{k}|$, $k \in {\rm I\!R}$, then all $k$ modes decay and the uniform state is linearly stable. It is easy to see that $\omega_{-}(k) \leq \omega_{+}(k)$. Thus, if $\omega_{+} > 0$ for some values of $k$, then such modes grow over time and will dominate the density modulation of the system. The onset of linear instability, i.e. when both $\omega(k)=0$ and $d\omega(k)/{dk}=0$ corresponds to solving $\textrm{det}(\textbf{E})=0$~\cite{dispersion_relation}, which is equivalent to the condition 
\begin{equation}\label{D_of_k}
\Delta(k)\equiv \left[1-\rho_{1}^0\hat{c}_{11}(k)\right]\left[1-\rho_{2}^0\hat{c}_{22}(k)\right]-\rho_{1}^0\rho_{2}^0\hat{c}_{12}^2(k)=0.
\end{equation}

It is worth mentioning that the values of the diffusion coefficients $D_1$ and $D_2$ do not influence the thermodynamic equilibrium state, i.e. the minimum of the free energy and therefore are not further discussed.

\begin{figure*}[t!]
    \centering
    \includegraphics[width=1\linewidth]{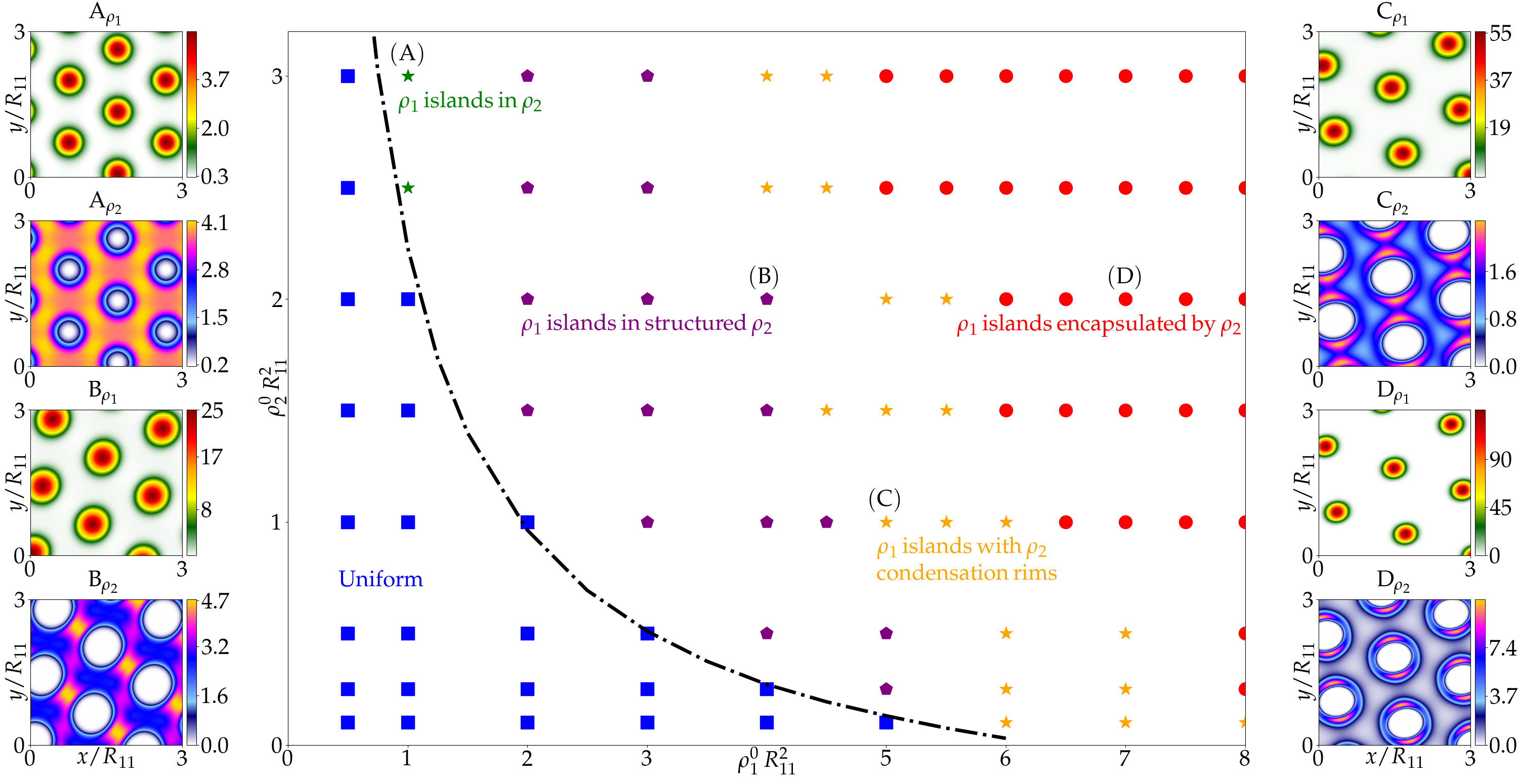}
    \caption{Phase diagram of the different states in the ($\rho^0_1, \rho^0_2$) plane corresponding to the model of Eq. (\ref{pairpot}) and parameters in Table \ref{table1}. The letters A, B, C, D inside the diagram identify the side panel visualizations for the two densities $\rho_1 R_{11}^2$ and $\rho_2 R_{11}^2$ at the labeled state point. 
    The linear instability line is shown by the dash-dotted line.}
    \label{pahse_diagram}
\end{figure*}
\subsection{The Ornstein-Zernike equation for binary mixtures}
The total correlation functions $h_{ij}(r)$ which are related to the radial distribution functions $g_{ij}(r)$ via $h_{ij}(r)=g_{ij}(r)-1$ can be obtained using the Ornstein-Zernike (OZ) equations. For a binary (isotropic) fluid mixture, the OZ equations are \cite{hansen_mcdonald}:
\begin{equation}
h_{ij}(r)=c_{ij}(r)+\sum_{k=1,2}\rho_{k}^0\int d\textbf{r}' c_{ik}(|\textbf{r}-\textbf{r}'|)h_{kj}(\textbf{r}').
\label{eq:OZ}
\end{equation}
These coupled equations must be solved in conjunction with the (exact) closure relations
\begin{equation}
c_{ij}(r)=-\beta\phi_{ij}(r)+h_{ij}(r)-\ln[1+h_{ij}(r)]+B_{ij}(r),
\label{eq:closure}
\end{equation}
where $B_{ij}(r)$ are the so-called bridge functions \cite{hansen_mcdonald}. For example, the HNC approximation consists of setting $B_{ij}(r)=0$ $\forall r$. The RPA approximation can be obtained by linearising the logarithm in Eq.~(\ref{eq:closure}) together with the HNC approximation. It is also worth mentioning that when modelling strongly repulsive and short-range interactions, the direct correlation functions can be obtained employing the Percus-Yevick approximation \cite{hansen_mcdonald}.
\subsection{The partial static structure factors}
The partial static structure factors, which are accessible via scattering experiments, are defined as \cite{hansen_mcdonald, dispersion_relation}:
\begin{equation}
\begin{split}
S_{11}(k)&=1+\rho_{1}^0\hat{h}_{11}(k);\\
S_{22}(k)&=1+\rho_{2}^0\hat{h}_{22}(k);\\
S_{12}(k)&=\sqrt{\rho_{1}^0\rho_{2}^0}\hat{h}_{12}(k),
\end{split}
\end{equation}
where the Fourier transform of the total correlation functions $\hat{h}_{ij}$ are
\begin{equation}
\hat{h}_{ij}(k)=\frac{N_{ij}(k)}{\Delta(k)},
\end{equation}
with the numerators $N_{ij}(k)$ given by
\begin{equation}
\begin{split}
N_{11}(k)&=\hat{c}_{11}(k)+\rho_{2}^0\left[\hat{c}_{12}^2(k)-\hat{c}_{11}(k)\hat{c}_{22}(k)\right];\\
N_{22}(k)&=\hat{c}_{22}(k)+\rho_{1}^0\left[\hat{c}_{12}^2(k)-\hat{c}_{11}(k)\hat{c}_{22}(k)\right];\\
N_{12}(k)&=\hat{c}_{12}(k).
\end{split}
\end{equation}
%
%
Note that in the case of a stable liquid, $\Delta(k)>0$ $\forall k$.

\section{Model for core-shell assemblies}\label{Sec:capsules}
\begin{figure}[b!]
    \centering
    \includegraphics[width=1\linewidth]{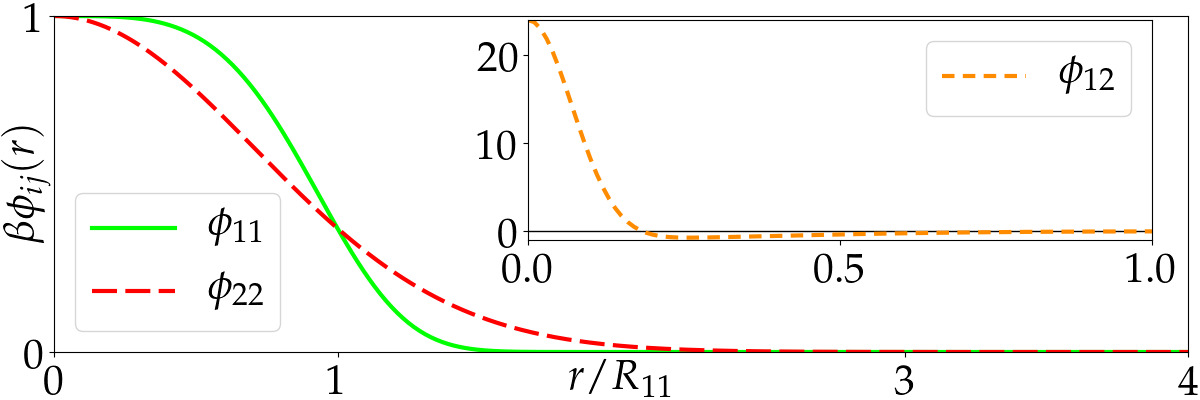}
    \caption{Different interaction potentials $\phi_{ij}$ with the parameters reported in Table \ref{table1}.}
    \label{potentials_1}
\end{figure}
For computational efficiency, we focus here on a two-dimensional (2D) model. We concentrate on a binary model that potentially self-assembles such that particle-like aggregates composed of a core and a corona, shielding the latter from bulk solution, can be expected. Practical examples of such assemblies are, e.g., surfactant micelles or block-copolymer particles. In aqueous environments, the core region is hydrophobic and the corona hydrophilic. To capture such compositional organization, we model the effective interactions of the different components as
%
\begin{align}\label{pairpot}
\phi_{\rm 11}(r)&=\varepsilon_{\rm 11}e^{-\left(r/R_{\rm 11}\right)^4}; \nonumber \\
\phi_{\rm 12}(r)&=\varepsilon_{\rm 12}^{-}e^{-\left(r/R_{\rm 12}^{-} \right)^2}+\varepsilon_{\rm 12}^{+}e^{-\left(r/R_{\rm 12}^{+} \right)^2};\\
\phi_{\rm 22}(r)&=\varepsilon_{\rm 22}e^{-\left(r/R_{\rm 22} \right)^2}, \nonumber
\end{align}
where $\beta\varepsilon_{\rm 12}^{-} < 0$ and $\beta\varepsilon_{\rm 12}^{+} > 0$, with the constraint $\vert \beta\varepsilon_{\rm 12}^{-}\vert < \vert \beta\varepsilon_{\rm 12}^{+}\vert$. Additionally, $R_{\rm 12}^{-} > R_{\rm 12}^{+}$. Together, these yield a short ranged repulsion and a long ranged attraction between the two species. Generally, $R_{11}$ and $R_{22}$ are a measure of the sizes of the two components (for polymeric systems these are comparable to the radius of gyration of the two species). The different variables chosen in this work for the model are reported in Table \ref{table1}, and the corresponding potentials $\phi_{ij}$ are shown in Fig.~\ref{potentials_1}.
\begin{table}[h!]
\centering
\begin{tabular}{| c | c | c | c | c | c | c | c | c |}\hline
$\beta\varepsilon_{\rm 11}$  & $\beta\varepsilon_{\rm 12}^{-}$ & $\beta\varepsilon_{\rm 12}^{+}$ & $\beta\varepsilon_{\rm 22}$  & $R_{\rm 11}$ & $R_{\rm 12}^{-}$ & $R_{\rm 12}^{+}$ & $R_{\rm 22}$\\  \hline\hline
1  & -1 & 25 & 1  & 1 & 0.5 & 0.1 & 1  \\   \hline
\end{tabular}\caption{Set of dimensionless parameters used in Eq.~(\ref{pairpot}) to model core-shell particles.}\label{table1}
\end{table}
Note that for the cross-interaction, the integrated strength $2\pi\int_0^{\infty} r \phi_{12}(r)dr=0$. The different potentials in Eq.~(\ref{pairpot}) have been chosen following the fact that in order to see the formation of clusters of purely repulsive particles, one needs in the model an intrinsic length scale that can eventually be unstable. Such property implies a symmetry breaking, and thus the formation of periodic (or aperiodic) patterns. In the current work, the interaction between particles of species 1 bear such a feature and is modelled using the so-called generalised exponential model of order $n$ (GEM-$n$) \cite{coslovich2012cluster, archer2014solidification, archer2016generation, archer2016crystallization, scacchi2018flow, caprini2019comparative, gotze2006structure, overduin2009phase, overduin2009clustering, camargo2010dynamics, camargo2011interfacial}. In this work $n=4$.  In order to keep the model as simple as possible, species 2 is modelled using the Gaussian core model (GCM), which does not belong to the class of $Q\pm$-interactions \cite{likos2007ultrasoft}, i.e. it does not lead to cluster formation in single-component mixtures (except for $\varepsilon\beta > 100$~\cite{louis2000mean}, irrelevant here). The GCM has been extensively used to model polymeric interactions \cite{bolhuis2001accurate, likos2001effective, likos1998star,louis2000can, gotze2004tunable, likos2006soft, lenz2012microscopically, mladek2006formation, gotze2006structure}. The second term in the cross-interaction $\phi_{12}$ models a short range repulsion, which is necessary to avoid overlap of particles of different species. However, in order to favor the formation of a shell surrounding the core of the particle, attraction between different species is also necessary. The latter is achieved via the first term in $\phi_{12}$. The range of the attractive and repulsive part of $\phi_{12}$ are intimately related to molecular details of the components. They can be controlled, e.g., by molecular weight, solvent composition, or changing  salt concentration or pH of the solution, depending on the nature of the species. It is worth mentioning that the results presented here are expected to be valid for a variety of interaction potentials which bear properties similar to those in the current model, Eq.~(\ref{pairpot}). The results can also be expected to hold for a large collection of parameters different from those reported in Table~\ref{table1}.
\subsection{Results}
\begin{figure}[h!]
    \centering
    \includegraphics[width=1\linewidth]{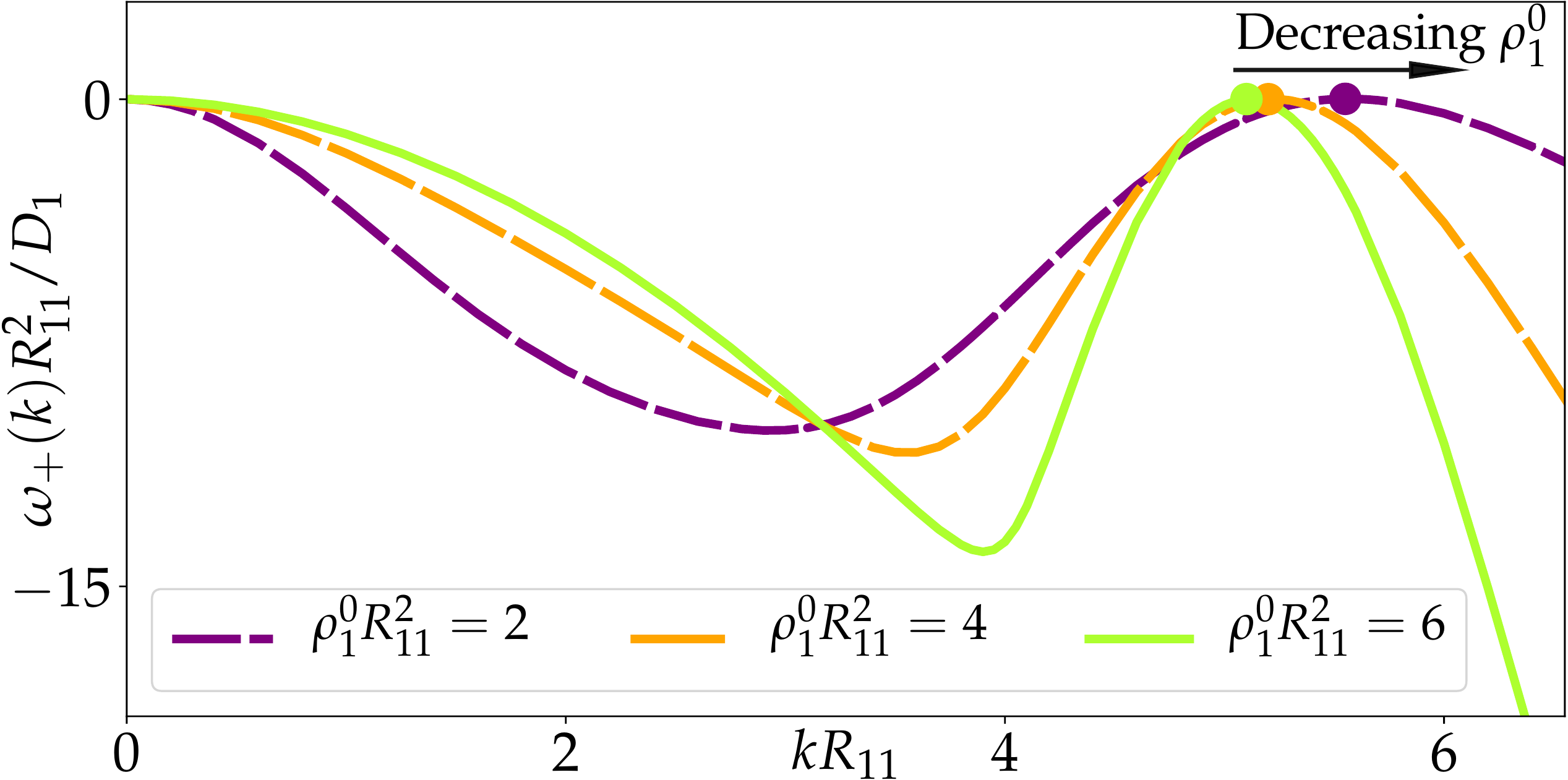}
    \caption{Different onsets of the linear instability, i.e. the marginal condition for density modulation in the system, as a function of $\rho_1^0$ (dash-dotted line in the central panel of Fig.~\ref{pahse_diagram}). The critical wave numbers $k_{\rm c}$, marked by circles, shift to higher values as $\rho_1^0$ decreases, which corresponds to the lattice constant of the expected periodic structure decreasing. Additionally, the peaks in the dispersion relations broaden as the bulk density of species 1 decreases. This indicates the selectivity of the wave number $k_c$ softens.}
    \label{dispersion_plot}
\end{figure}
 We explore the density-density plane ($\rho^0_1, \rho^0_2$) to resolve the equilibrium self-assembly phases adopted by a system following the interaction parameters reported in Table \ref{table1}. Here $\rho^0_1$ and $ \rho^0_2$ are the bulk number densities of species 1 and species 2, respectively. The resulting assembly phase diagram and the four different assembly configurations adopted by the system in different composition regions are shown in Fig.~\ref{pahse_diagram}. For each state point, we report the density of both species 1 and species 2, labelled accordingly. The phase diagram indicates that the system adopts a homogenous mixture corresponding to uniform solution of species 1 and 2 at low concentrations of $\rho_1$ while an increase of $\rho_1$ leads to an onset of species 1 condensation into islands that exhibit hexagonal symmetry. This occurs at a concentration that is dependent on $\rho_2^0$. The data of Fig.~\ref{pahse_diagram} indicate that the system has only one phase transition. The transition happens when passing from the homogeneous state ({\it uniform}) to any state with hexagonal symmetry in species 1 assembly pattern. The presence of a single phase transition is expected, since the system has only one intrinsic length scale. 

While species 1 assembles into condensation islands with hexagonal symmetry in spatial locations throughout the structure-forming part of the phase diagram, species 2 shows interesting configurational changes when exploring different concentrations. In the first two of these, species 2 adopts a distribution where it avoids the locations where species 1 is enriched as clusters. A region with very low density, i.e. a depletion region of $\rho_2$ forms around the $\rho_1$ islands. In the {\it $\rho_1$ islands in $\rho_2$} states of Fig. \ref{pahse_diagram}, most of $\rho_2$ forms a uniform background solution with density very close to the bulk value $\rho_2^0$ in the regions where $\rho_1$ is absent.  An example of this configuration is shown in Fig.~\ref{pahse_diagram}, panels ${\rm A}_{\rho_1}$ and ${\rm A}_{\rho_2}$. Such assembly corresponds to, for example, multicomponent polymer solutions in which one species forms polymer particles and another remains soluble, yet immiscible with the former species. One can consider this also as two immiscible liquids, with $\rho_1$ forming droplets surrounded by the phase in which $\rho_2$ spreads uniformly to form a continuous phase. Such phase separation occurs, for example in intracellular liquid-liquid phase separation or coacervation of macromolecule solutions. 

In states labeled as {\it $\rho_1$ islands in structured $\rho_2$} in Fig.~\ref{pahse_diagram}, similar holes in $\rho_2$ around the $\rho_1$ condensation sites characterize the density distribution. However, $\rho_2$ distribution shows an underlying periodic structure in the regions where particles 2 are in the liquid state (see panels ${\rm B}_{\rho_1}$ and ${\rm B}_{\rho_2}$ of Fig.~\ref{pahse_diagram}). This represents species 1 driving correlations into the structure formation of species 2, despite species 2 retaining partial liquid character. Transition from the uniform background species 2 to this correlated species 2 distribution is an indication of increasing effective attraction between the assembled species 1 islands via the species 2 condensation. This would show in practical systems as increasing solution viscosity. 

The {\it $\rho_1$ islands with $\rho_2$ condensation rims} state points correspond to configurations in which species 2 gather around the peaks of $\rho_1$ defining the primary hexagonal structure. Away from these condensation islands however, the density of both $\rho_1$ and $\rho_2$ is depleted. These state points correspond to a partial (or soft) encapsulation of species 1 by a coating of species 2. An example of the latter configuration is shown in Fig.~\ref{pahse_diagram}, panels ${\rm C}_{\rho_1}$ and ${\rm C}_{\rho_2}$. Such assembly occurs, for example, in amphiphile mediated solubilization or dispersion under conditions where part of the coating species remains soluble. This assembly also corresponds to part of the species 2 assembling into micellar or core-shell particles, part remaining soluble.

Finally, the region labeled {\it $\rho_1$ islands encapsulated by $\rho_2$} correspond to structures where species 2 surrounds (almost) completely the clusters of species 1. This leads to a strong encapsulation configuration corresponding to micellar or core-shell type well-defined layered self-assembly structures in e.g. surfactant and polymer systems. This configuration is shown in panels ${\rm D}_{\rho_1}$ and ${\rm D}_{\rho_2}$ of Fig.~\ref{pahse_diagram}. The asymmetry in the shells is a result of the relatively strong and long-ranged repulsion between particles of species 2 via $\phi_{22}(r)$. The asymmetry can be reduced by diminishing the strength and range of this repulsion (see Appendix). The instability line obtained from the dispersion relation $\omega_{\pm}(k)$ is reported in the central panel of Fig.~\ref{pahse_diagram}. Left of this line the system is expected to be linearly stable which leads to the equilibrium density distribution being homogeneous. At right of this line, however, the system is expected to form patterns with lattice spacing corresponding to $\approx 2\pi/k_{\rm c}$. Here $k_{\rm c}$ is the wave number at which the system is marginally unstable. This corresponds to $\omega_{+}(k_{\rm c})=0$ and ${d\omega_{+}(k_{\rm c})}/{dk}=0$. Careful examination of the critical wave number $k_{\rm c}$ value for different pairs of bulk densities $\rho_1^0$ and $\rho_2^0$ reveals a monotonic shift such that moving along the instability line from high to low values of $\rho_1^0$,  $k_{\rm c}$ increases. This means also that the lattice constant of the underlying hexagonal structure decreases. Additionally, the peak in the dispersion relation broadens, which translates into a less selective instability. The latter result is reported in Fig.~\ref{dispersion_plot}. 

It is worth mentioning that at temperatures relevant here, two dimensional monodisperse systems with only hexagonal structures of lattice spacing $\approx R_{11}$, i.e. approximately constant lattice spacing for all densities, are expected~\cite{archer2015soft}. However, this is not true in three dimensions, where a series of isostructural phase transitions are expected at low temperatures~\cite{zhang2010reentrant,wilding2014demixing}.

\section{Model for mixed particles}\label{Sec:core_shell}
Exchanging the short-ranged repulsion and the long-ranged attraction with a short-ranged attraction and a long-ranged repulsion enables obtaining density distributions which are commensurate with mixed assembled particles. This means $\beta\varepsilon_{\rm 12}^{-} < 0$, $\beta\varepsilon_{\rm 12}^{+} > 0$, with the constraint $\vert \beta\varepsilon_{\rm 12}^{-}\vert > \vert \beta\varepsilon_{\rm 12}^{+}\vert$, and also $R_{\rm 12}^{-} < R_{\rm 12}^{+}$. Monodisperse suspensions interacting via a different short range attraction and long ranged repulsion (SALR) potential in two dimensions exhibit fascinating microphase separation~\cite{archer2008two}. Noteworthy is also the interesting study of phase behavior of a SARL fluid in three dimensions~\cite{archer2007phase}. Following Section \ref{Sec:capsules}, the effective pair potentials used in this model are described by Eq.~(\ref{pairpot}), however, the coefficients differ from Sec.~\ref{Sec:capsules} and are reported in Table \ref{table2}. To control the size and composition of the particles we vary the range of the attractive and repulsive contributions in $\phi_{12}$, i.e. $R_{12}^{-}$ and $R_{12}^{+}$, respectively, keeping $R_{\rm 12}^{-} < R_{\rm 12}^{+}$. Note that the integrated strength of the cross-interaction is positive for all five cases discussed here, i.e. $2\pi\int_0^{\infty} r \phi_{12}(r)dr>0$.
\begin{table}[h!]
\centering
\begin{tabular}{| c | c | c | c | c | c | c |}\hline
$\beta\varepsilon_{\rm 11}$  & $\beta\varepsilon_{\rm 12}^{-}$ & $\beta\varepsilon_{\rm 12}^{+}$ & $\beta\varepsilon_{\rm 22}$  & $R_{\rm 11}$ &  $R_{\rm 22}$\\  \hline\hline
1  & -1 & 0.5 & 1 & 1 & 1  \\   \hline
\end{tabular}\caption{Set of dimensionless parameters used in Eq.~(\ref{pairpot}) to model mixed particles.}\label{table2}
\end{table}
\subsection{Results}
\begin{figure}[h!]
    \centering
    \includegraphics[width=1\linewidth]{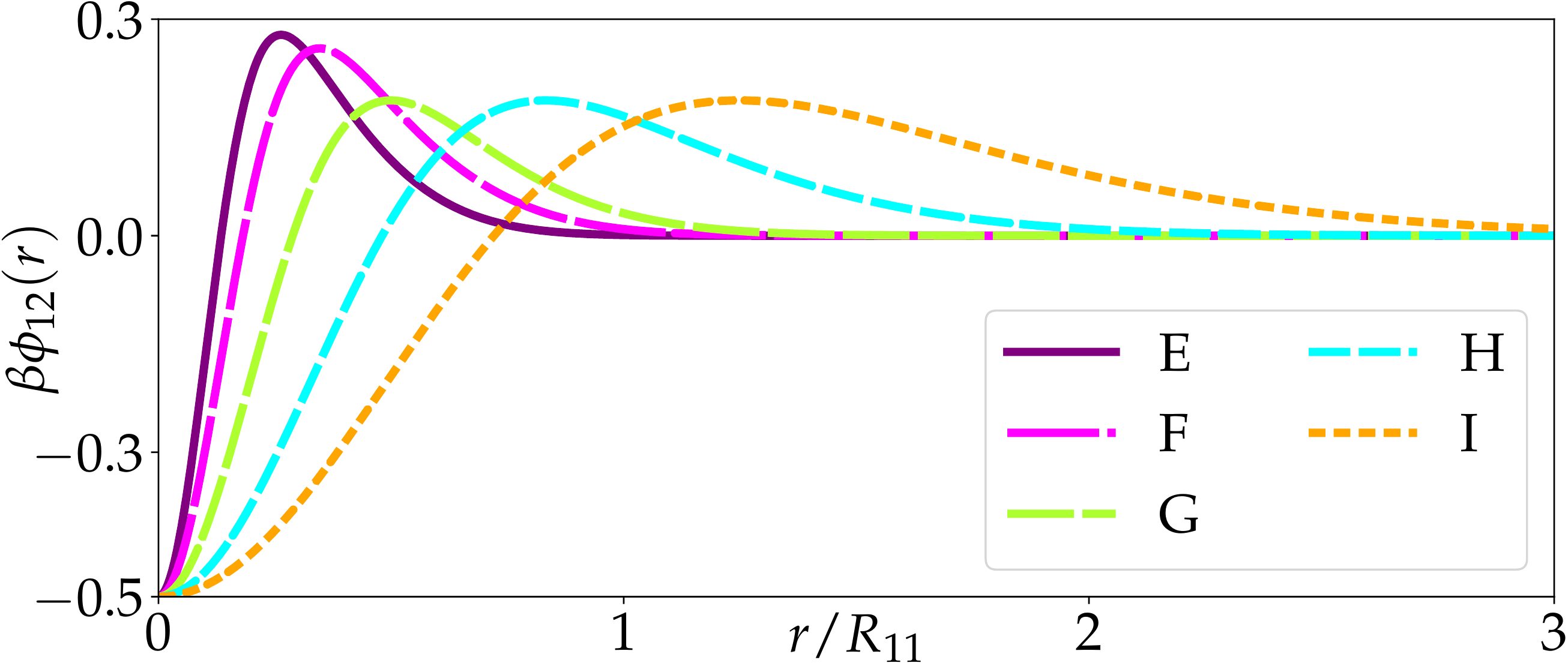}
    \caption{Cross-interaction potentials $\phi_{12}$ following the parameters reported in Table \ref{table2} for varying values of $R_{12}^{-}$ and $R_{12}^{+}$ in Table \ref{table3}.}
    \label{potentials}
\end{figure}
\begin{figure*}[t!]
    \centering
    \includegraphics[width=1\linewidth]{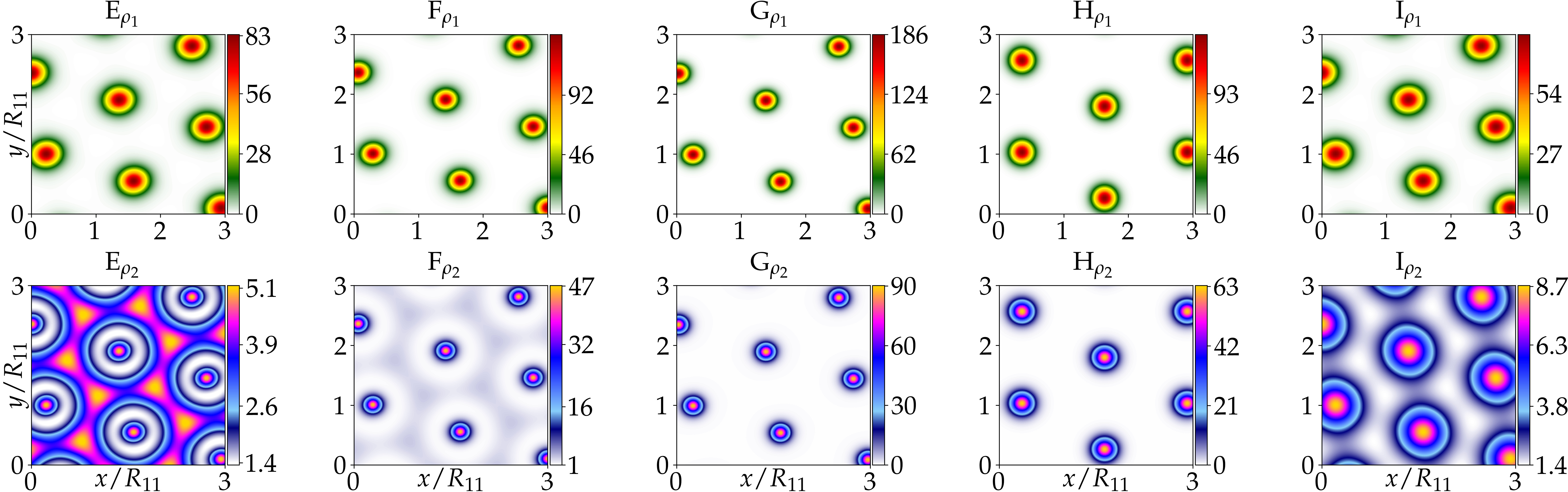}
    \caption{Average density distribution $\rho_i R_{11}^2$ for $i=1$ (top row) and for $i=2$ (bottom row) for the set of parameters corresponding to configurations E, F, G, H and I. The parameters used for modelling $\phi_{12}$ are described in Table \ref{table2} and the varying values of $R_{12}^{-}$ and $R_{12}^{+}$ are reported in Table \ref{table3}.}
    \label{core_corona}
\end{figure*}
We set the average densities of the two components of the system to $\rho_1^0R_{11}^2=6.5$ and $\rho_2^0R_{11}^2=3$, respectively, and study the equilibrium configurations obtained using the DFT described in Section \ref{sec:DFT} as a function of different cross interaction potentials. 
\begin{table}[h!]
\centering
\begin{tabular}{| c || c | c | c | c | c | }\hline
Model & E & F & G & H  & I \\  \hline\hline
$R_{12}^{-}$ & 0.15 & 0.2 & 0.3 & 0.5 & 0.75  \\   \hline
$R_{12}^{+}$ & 0.4 & 0.5 & 0.6 & 1 & 1.5    \\   \hline
\end{tabular}\caption{Range of the attractive and repulsive contribution in the cross particles interaction potential $\phi_{12}(r)$ leading to mixed particles with different density gradients.}\label{table3}
\end{table}
The form of the inter-species interaction potentials described by the parameters in Table \ref{table3} are shown in Fig.~\ref{potentials}. 

Figure \ref{core_corona} shows the resulting equilibrium density distributions of both species with species 1 as the top row and species 2 as the bottom row. As before, species 1 assembles into condensation islands with hexagonal spacing throughout the examined parameter range. Only minor changes in condensation peak distribution height and width can be observed. However, rather interesting assembly changes occur to species 2. In case E, the very short attraction causes a relatively small increment of $\rho_2$ around the locations occupied by species 1. The repulsive part of $\phi_{12}$ however causes a depleted region around the latter peaks. Here $\rho_2$ forms a thin, skin-like, layer around the species 1 condensation island and is separated from bulk by a depletion region. Beyond the depletion region, species 2 adopt an almost liquid-like assembly but showing preferred density locations as triangles formed between the $\rho_1$ neighboring peaks. This assembly pattern means that also for the species 2, hexagonal order with lattice parameter differing from that of species 1 ordering emerges. 

In case F, the total attraction strength between the two components is increased. The attraction is strong enough to pull almost all species 2 particles to the sites enriched in species 1. The clusters of species 2 are very narrow for this set of parameters, and actually now species 1 forms the coating layer. However, no clear core phase exists and the two species remain mixed in the island center regions, however with strong concentration gradients. A low density honeycomb-like region of $\rho_2$ is visible far from these condensation peaks indicating some of species 2 remaining soluble outside species 1 islands. 

Similar to case F, also in setup G, $\rho_2$ exhibits sharp and narrow peaks at the locations of particles 1. The width of the $\rho_2$ peaks is practically the same as those of $\rho_1$. This means that both species occupy the same region and readily mix. Moving to cases H and I, the widths of both $\rho_1$ and $\rho_2$ peaks increase. Eventually, the peaks of $\rho_2$ become larger than those of species 1 leading to a case where species 1 can be found mostly in the centre of the particles. Even in this case I, also $\rho_2$ is enriched at the same location meaning that both density and condensation have a gradient in the island. 

\begin{figure}[h!]
    \centering
    \includegraphics[width=1\linewidth]{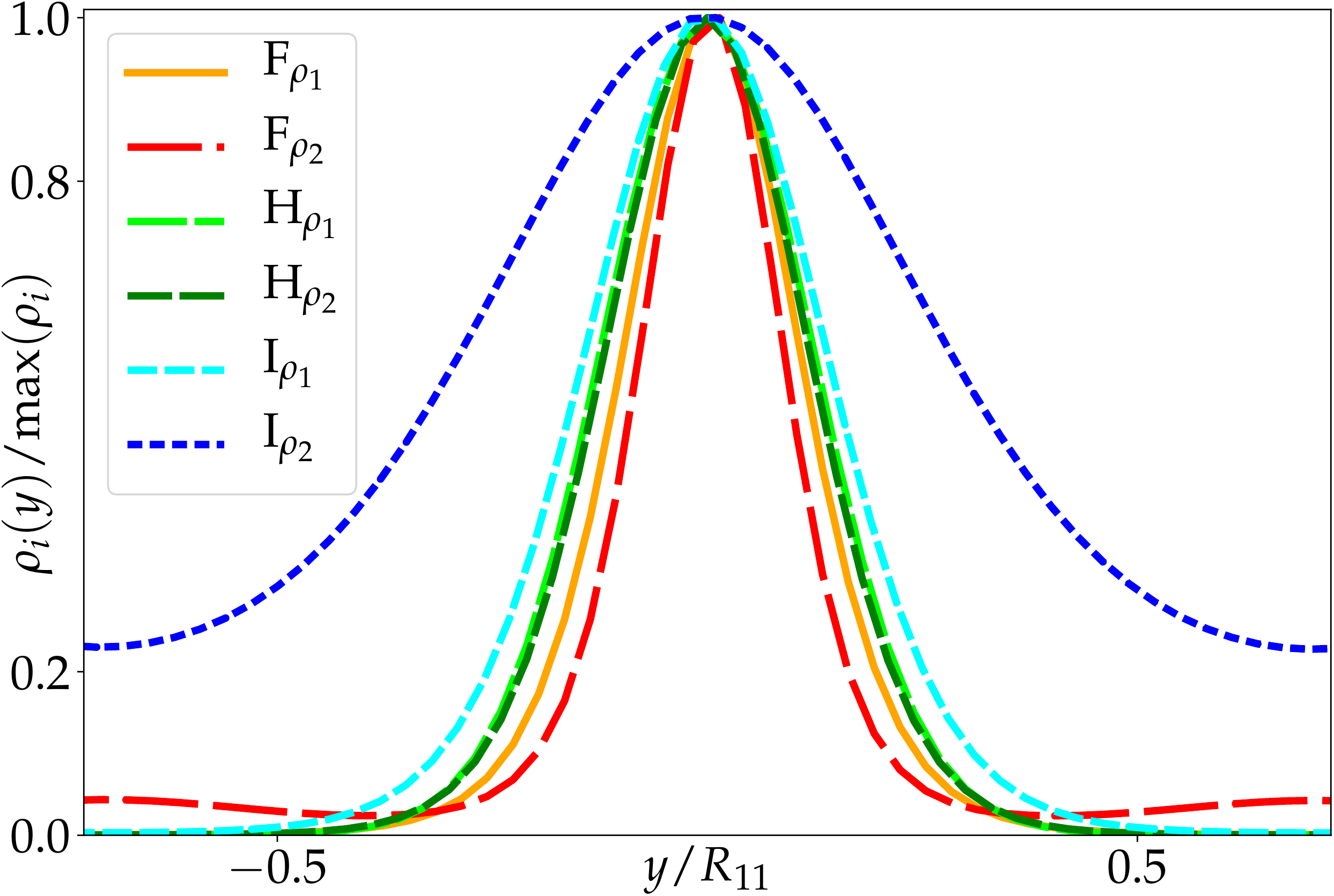}
    \caption{One-dimensional density distributions $\rho_1(y)$ and $\rho_2(y)$ of a single condensation peak for cases F, H and I. The density is normalised by its maximal value. Minor peak size changes are found in $\rho_1$ for the different cases. However, $\rho_2$ peaks change significantly. In case F, species 2 occupies mainly the center of the particle. In case H, the two species are very well mixed. In case I, species 1 occupies the center of the aggregate.}
    \label{cut}
\end{figure}

To summarize the assembly findings, Figure \ref{cut} shows the 1D density profiles of the islands for states F, H and I. The data are normalised by the maximal values of the distributions and centered at the maximum of the peak. The figure visualizes how to control the size of the mixed particles, as well as guidelines to engineering the density gradient of the two species mixing inside the particle, but also the overall particle density profile. At a practical level, such concentration gradients mean variations in the local chemical environment. This has a major significance for applications. For example, hydrophilicity-hydrophobicity gradients control drug species partitioning in the carrier and in-out diffusion, which control the binding and release of carried species. 
\section{Summary and Conclusions}
Using classical density functional theory, we have presented a general framework for particle systems with competing molecular interactions. We have shown how this can lead to the spontaneous formation of structurally complex self-assembly matter such as core-shell structures and well-defined assemblies where the component species mix, yet granting control on mixing properties. We have identified the key intermolecular interactions features responsible for these complex self-assembling structures and demonstrated how the assembly and species mixing can be controlled. To this end, linear stability analysis was used as a guiding principle.

This insight can be employed to better understand multi-component self-assembly systems that emerge in, e.g., cellular function and complex biological structure formation. It also paves the way for engineering advanced complex materials solutions for, e.g., drug delivery including anticancer therapy approaches that often rely on sequential delivery of multiple drugs, see e.g. \cite{Aw2013,Wu2018}. 

In this work, we applied DFT to two different cases. First, we mapped out a phase diagram where one species occupies the core of the self-assembled particles and the second species surrounds the latter either as a freely solvated species or a corona layer depending on the concentrations of the two components. This assembly response, decribed in Sec.~\ref{Sec:capsules}, can be considered to correspond to spontaneous formation of droplets (for example, biomolecular condensation), polymer core-shell particles or surfactant micelles. Such systems are ubiquitous in biological systems but also chemical engineering such as paints, cosmetics, and pharmaceutical formulations. 

We also demonstrated that via the effective interactions between the species, it is possible to carefully control not only the size of the self-assembling particles but also their internal composition gradient and mixing of the species inside the particles. Furthermore, the findings give access to controlling overall density distribution. These findings are summarized in Sec.~\ref{Sec:core_shell}. A handle to self-assembling particle internal composition has applications in molecular partitioning (hydrophobicity gradients), diffusion control, chemical reactions engineering but also in controlling solubilization and dispersion. 

We note that the present theory is not restricted to (ultra) soft interaction potentials that were the main focus in this work. In fact, using different approximations for the free energy functionals, the theory developed here can be applied to a wide range of systems, including, e.g., colloids and nanoparticles.
Additionally, one should note that the theoretical approach presented here can also be applied to systems exhibiting multiple length scales in their self-assembled configurations~\cite{scacchi2021self}. Examples of such systems that constitute coexisting structures with different length scales include, e.g., multi-core micelles \cite{wang2011discovering, chen2013formation, duxin2005cadmium, iatridi2011ph, ueda2011unicore, chen2012formation}, coacervate droplets in biocondensates (liquid-liquid phase-separation in biological systems) \cite{lu2020multiphase} and multiple emulsions \cite{sheth2020multiple}.
\section{Acknowledgment}
This work was supported by Academy of Finland grants No. 309324 (M.S.) and Nos. 307806 and 312298 (T.A-N.). T.A-N. has also been supported by a Technology Industries of Finland Centennial Foundation TT2020 grant. We are grateful for the support by FinnCERES Materials Bioeconomy Ecosystem. Computational resources by CSC IT Centre for Finland and RAMI -- RawMatters Finland Infrastructure are also gratefully acknowledged.
\section{Author contributions}
A.S. conceived the work, executed the theoretical calculations and wrote the first draft of the manuscript. M.S. and T.A-N. supervised the research. All authors contributed to writing.
\section{Data availability statement}
The data that support the findings of this study are available within the article. Additional data are available upon request.
\appendix
\section{Asymmetry in capsule-like structures}
\begin{figure}[h!]
    \centering
    \includegraphics[width=1\linewidth]{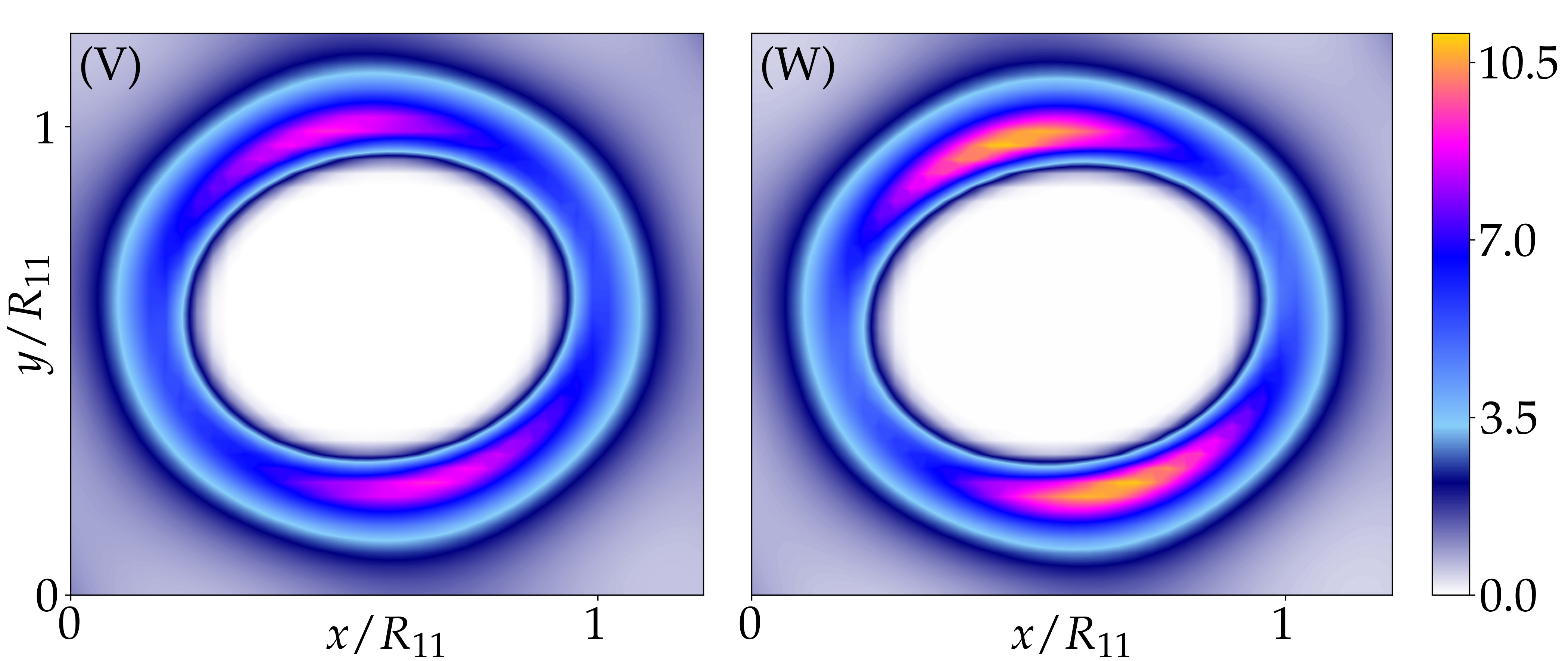}
    \caption{Asymmetry dependence on $R_{22}$ interaction parameter. Panel V shows $\rho_2(\textbf{r})$ for $R_{22}=0.24$ and panel W for $R_{22}=1$. All other parameters are the same as in case D of Fig.~\ref{pahse_diagram}. The asymmetry in the density distribution is visibly higher in panel W which corresponds to a longer-ranged $\phi_{22}$. The maximal density value in W is roughly $20$~\% higher than in V.}
    \label{asymmetry}
\end{figure}

The asymmetries in the equilibrium densities reported in Fig.~\ref{pahse_diagram} result from the relatively strong and long-ranged repulsion between particles of species 2 that compete with all other energy contributions in the system. This asymmetry can be diminished by reducing the integrated strength of $\phi_{22}$. This can be done by changing either the strength or the range of $\phi_{22}$, i.e. by decreasing $\beta\epsilon_{22}$ or $R_{22}$ (or both). To demonstrate this, we keep the repulsion strength between particles of species 2 at $\beta\varepsilon_{22}=1$, but change the range of the interaction by setting $R_{22}=0.24$. Fig.~\ref{asymmetry} presents a comparison of the the density distribution of species 2 corresponding to $R_{22}=0.24$ (panel V) and $R_{22}=1$ (panel W). Panel W corresponds to the case reported in panel ${\rm D}_{\rho_2}$ of Fig.~\ref{pahse_diagram}.
\newpage
\bibliographystyle{ieeetr}
\bibliography{main.bib}

\end{document}